\PassOptionsToPackage{hidelinks}{hyperref}
\PassOptionsToPackage{hyphens}{url}
\documentclass[preprint,authoryear,12pt]{elsarticle}

\usepackage{lmodern}
\usepackage[T1]{fontenc}
\usepackage{amsmath,amssymb}
\usepackage{amsthm}
\usepackage{booktabs}
\usepackage{array}
\usepackage{microtype}
\usepackage{url}

\graphicspath{{figures/}}

\newtheorem{proposition}{Proposition}
\makeatletter
\def\ps@pprintTitle{%
  \let\@oddhead\@empty
  \let\@evenhead\@empty
  \def\@oddfoot{\reset@font\hfil\thepage\hfil}%
  \let\@evenfoot\@oddfoot}
\makeatother

\begin{document}

\begin{frontmatter}

\title{Seasonal false alarms in customer churn and decline early-warning
systems: adjacent-window labels confound seasonality with decline, and a
year-over-year correction}

\author[fn]{Md Rezwanul Islam\corref{cor1}}
\ead{rezwanul.islam@fieldnation.com}
\author[fn]{Wael Mohammed}
\ead{wael.mohammed@fieldnation.com}
\ead[url]{https://orcid.org/0009-0005-4211-8976}
\affiliation[fn]{organization={Field Nation LLC},
                 city={Minneapolis}, state={MN}, country={USA}}
\cortext[cor1]{Corresponding author. ORCID:
\texttt{https://orcid.org/0009-0002-4100-3796}.}

\begin{abstract}
Customer decline early-warning systems feed account-manager action
lists, and every flagged account consumes intervention capacity. In a
deployed business-to-business marketplace system, one action-list slot
in three went to flags that dissolve under a seasonally aligned
label. The standard target in non-contractual churn prediction
compares an entity's next $k$ months of activity with its trailing $k$
months. The two windows cover different calendar months, so for seasonal
entities the threshold-ratio construction confounds seasonality with
decline, and the event rate depends on the label's anchor calendar
month. We formalize the mechanism and measure it on three
public panels and the production panel (the public arms pre-specified).
Of the adjacent-window decay events, 37--69\% on the public panels and
28--50\% in production have no counterpart under a seasonally aligned
definition. Pooling anchors, the standard remedy, balances the training
mixture but corrects no individual label. Measured alternatives
repair the curve only partially or change the detection horizon.
Aligning the baseline to the same $k$ calendar months one year prior ---
advice practitioners already state, here formalized, measured, and
costed --- flattens the curve at the source. With the classifier held
fixed, production holdout ROC-AUC rises from 0.767 to 0.864 for decline
(different targets; the gap closes on synthetic ground truth and a
production hindsight referee). The served action list
shrinks by a third, 119 to 79 accounts. The measured price is extra
history, a blind spot to decline-then-stabilization, and a stricter cut
under sustained growth.
\end{abstract}

\begin{keyword}
Decision support systems \sep churn prediction \sep customer base analysis \sep
label construction \sep seasonality
\end{keyword}

\end{frontmatter}

\section{Introduction}
\label{sec:intro}

In a deployed buyer-health system at a business-to-business service
marketplace, an account at its all-time spend high was flagged as
shrinking. The flag consumed an account-manager slot that a genuine
decliner needed. The audit that followed traced the false alarm not to
the model but to the label the model had learned from. Roughly one
served flag in three dissolved under a seasonally aligned relabeling.
In non-contractual settings there is no cancellation event to
observe, so decline targets are built from the transaction history
itself. The usual construction compares two adjacent windows of the
same entity's activity. It sums spend or purchase frequency over the
$k$ months before an anchor date and over the $k$ months after it. The
entity is labeled as declining when the forward sum falls below a
threshold fraction of the trailing sum. The two-window
design goes back to \citet{buckinx2005}. The within-entity
threshold-ratio family \citep{migueis2012} runs through the decline
targets that followed. Window design --- cutoff times, observation
and outcome lengths --- is standard material in deployed tooling and
label-engineering libraries \citep{msdyn365churn,composeml}.
None of these sources mentions the calendar placement of the windows.
Ratio labels are one of two major label families in non-contractual
settings; the other declares churn from inactivity alone. Seasonal
quiet can masquerade as churn there too. This paper analyzes and
corrects the threshold-ratio form only.

The construction has a defect: its two windows cover \emph{different
calendar months}. For a seasonal entity the two windows sit at different
points of the seasonal cycle. The label therefore fires on seasonal
descent and heals on seasonal ascent, whether or not the entity's
trajectory has changed. The event rate then becomes a function of the
calendar month at which the labels are anchored. A model trained on such
labels is partly a seasonality detector. The accounts it flags on any
scoring date include entities simply entering their low season. The
symptom is not unknown: one peer-reviewed study plots its defection rate
by anchor slice and reads the swings as ``the inherent seasonality of
customer defection'' \citep{georgiou2022}. What the literature does not
contain, to our knowledge, is the attribution. No source demonstrates
that the construction manufactures the pattern. None measures how much
of the labeled event mass dissolves under a seasonally aligned
definition, and none corrects the label itself. Those three steps are
this paper.

The measurement came out of production. The system of the opening
paragraph was a pair of gradient-boosted classifiers predicting
six-month spend decay and growth from an adjacent-window target. Across
its 10{,}426 buyer--month training observations, the decay-event rate
ranged from 2.4\% at April anchors to 9.3\% at October anchors. About
half of the decay events dissolved under a seasonally aligned
definition. The defect is a property of the construction, not of one
firm's data. Pre-specified replications on three public panels (M5
retail unit sales, Australian regional tourism, UCI Online Retail~II
per-customer wholesale spend) reproduce the artifact wherever decline
events occur at all. Rate ratios run 2.1 to 4.5, and
aligned-disagreement shares run 37\% to 69\%. A synthetic experiment
isolates the mechanism. When seasonal amplitude is the only quantity
manipulated, the false decay rate climbs from below 0.01\% to 27.3\%. A
seasonally aligned label stays at the noise floor.

The correction is the comparison analysts already use for seasonal
business numbers: judge a period against the same period one year
earlier. Official statistics treats that reading as absorbing stable
seasonal effects \citep{ess2024}. What the peer-reviewed literature
lacks, to our knowledge, is the label form. That form is a supervised
decline label, defined by the forward window's shortfall against the
\emph{same $k$ calendar months one year prior}. It also lacks any quantification
of the artifactual share or of what alignment changes. Both are
contributed here. The aligned label flattens
the anchor-month rate curve on every real-data panel we test. It is not free.
It needs extra history and a materiality fence. It goes quiet on
decline followed by stabilization, although it in fact catches
multi-year decliners more often than the adjacent label
(Section~\ref{sec:results-decomp}). Under sustained strong growth it
applies an effectively stricter cut (Section~\ref{sec:results-robust}).
These costs are measured alongside the benefits. It also differs from
the remedies on record. Those pool anchor slices or lengthen windows,
and they leave each example's wrong stamp in place
(Sections~\ref{sec:results-pool} and~\ref{sec:results-remedies}).
Finally, two controlled comparisons check that label quality turns into
model quality. The production classifier was retrained with features,
learner, and temporal holdout held fixed; only the target changed.
Holdout ROC-AUC rises from 0.767 to 0.864 for decay and from 0.736 to
0.857 for growth. Those two figures score different targets, so a
skeptic will rightly ask for a common referee. On synthetic data, the
same learner trained on the aligned label detects genuine declines in
progress (ROC 0.78). The adjacent-trained learner ranks them below
chance (0.44), and calendar-month features repair nothing.
Section~\ref{sec:results-robust} repeats the check against a hindsight
referee on the production panel itself.

The contribution is fourfold: (i) attribution and measurement of an
anchor-month event-rate artifact intrinsic to the standard
adjacent-window decline label, on one production and three public
panels; (ii) a mechanism account with a short analytic result --- the
twelve anchor-month label ratios of a seasonal entity have geometric
mean one, so any profile with unequal window sums places some anchors
below parity; (iii) a label-level correction,
year-over-year alignment of the outcome comparison, measured against the
standard alternatives with its benefits and its costs; and (iv) the
measured decision-level impact in a deployed system --- with the
classifier held fixed, relabeling raised holdout ROC-AUC by 0.10 for
decay and 0.12 for growth and shrank the served action list by a third,
a capacity effect Section~\ref{sec:practice} bounds in
intervention-cost units. The paper
thereby adds a case to a broader body of evidence. The routine parts of
supervised pipelines --- evaluation splits \citep{bergmeir2012},
information leakage \citep{kapoor2023}, class composition
\citep{veganzones2018} --- can matter more than the modelling choices
layered on top. Here the
routine part is the target definition itself.

\section{Related work}
\label{sec:related}

\paragraph{The adjacent-window canon}
\citet{buckinx2005} define partial defection by re-evaluating two
population-benchmarked loyalty conditions in the five months that follow
the five months used for the features. In their data the baseline runs
April--August and the outcome September--January: the outcome contains
Christmas, the baseline does not. The single reported event rate
(25.15\%) belongs to that one calendar placement. With ten
months of data and one snapshot, the dependence could not have been
observed even in principle. The same two-window shape runs through
consumer and business-to-business churn prediction wherever the target
is constructed from transaction windows
\citep{glady2009,tamaddoni2014}. Benchmark suites in
largely contractual settings \citep{verbeke2012,decaigny2018} typically
observe a recorded cancellation event rather than a window-constructed
target. To that extent they sit outside the confound studied here.
\citet{migueis2012} give the threshold-ratio form its cleanest
statement: partial churn when every quarter after a reference quarter
stays below 40\% of that quarter's spend. That is a within-customer
ratio against a \emph{frozen} reference quarter rather than a sliding
trailing window, in a full text that never mentions seasonality. This
paper analyzes the family's adjacent-window member, whose baseline
slides with the anchor. \citet{neslin2006} fix observation and outcome
windows with no calendar dimension. Where window design has been
studied directly, the question has been length, not calendar placement
\citep{ballings2012,han2026}. Label-definition choices are
known to move churn results without the calendar being examined
\citep{bugajev2022,mirkovic2022}. \citet{coussement2017} report that
data-preparation choices alone improve churn AUC by up to 14.5\%. The
same journal's churn line has extended the pipeline on other axes:
life-event prediction from fine-grained transaction data
\citep{decaigny2020}, and interpretable and hybrid classifier designs
\citep{debock2021,decaigny2024}. The label's calendar anchor is the
preparation choice this literature has not yet measured. Among
operational-research neighbours, \citet{clemente2014} make the
partial-defection definition the object of a profit-driven search over
variable subsets and window lengths, anchored at one fixed date.
\citet{mahajan2020} classify business-to-business revenue changes
between fixed epochs. The field's flagship retention review corrects
decline signals for one timing distortion: recency judged against the
entity's own inter-purchase cadence \citep{ascarza2018}. That
correction is cycle-relative, not season-relative. Closest in time,
\citet{ulrich2026} audits return-incidence labels at 18--24-month
horizons and remarks that whole-year horizons ``integrate the outcome
window over the seasonal cycle exactly''. At short horizons nothing
integrates out. Return-incidence labels also lack the trailing
comparator through which the confound studied here enters. None of
this implies the lineage was careless. It means a defect of the shared
construction went unmeasured.

\paragraph{Slice pooling and practitioner advice}
The nearest published work in our setting is \citet{gattermann2021}.
They show in a business-to-business non-contractual panel that training
on a single time slice overfits its ``distinctive situation'' and that
stacking slices improves churn prediction. They name event rarity,
trends and seasonality, and situation-specific patterns as the problem.
Their remedy operates on the composition of the training set. The
multiple-snapshot idea is older \citep{gurali2014}. Credit scoring
gives the matching scorecard advice \citep{siddiqi2006}. An industrial
patent recommends month-shifted overlapping training sets partly
because doing so ``weakens seasonal effects''
\citep{accenture2014patent}. \citet{georgiou2022} extend multi-slicing
for e-commerce defection with a two-window, within-customer label (an
interpurchase-gap threshold rather than an activity ratio, so
Proposition~\ref{prop:mechanism} does not literally cover their form).
They plot their defection rate by anchor slice and read the swings as
``the inherent seasonality of customer defection''. Nothing in their
treatment traces the swings to the label, quantifies an artifactual
share, or changes the definition. All of these treat the calendar as a
sampling problem. Each example keeps whatever stamp its calendar
position gave it. Practitioner guidance, for its part, already states
the aligned comparison as advice: judge a seasonal customer against the
same season of the prior year. Year-over-year reading of business
metrics is folklore. What we found stated nowhere in full, practitioner
or academic, is the audit itself. No source traces the anchor-month
curve to the construction. None quantifies the artifactual share of its
events, and none follows the changed label into a trained model. The
contribution is therefore not the idea of year-over-year comparison but
the aligned label form and the measurement of what it changes.

\paragraph{Latent-attrition models}
The probabilistic customer-base lineage \citep{schmittlein1987,fader2009}
treats churn as a latent state rather than a constructed label.
(\citet{wubben2008} showed simple heuristics rival these models, one
reason discrete labels persist in practice.) Its modern branches
handle seasonality inside the model. \citet{bachmann2021} put
time-varying contextual factors on purchase and attrition rates, with
the finding that seasonality moves purchases while showing no
significant effect on attrition. \citet{wunderlich2022} model seasonal
purchase-level behavior with dropout. This lane contains the
closest earlier statement of our mechanism that we could find anywhere.
\citet[p.~30]{wunderlich2015} observes that a three-month purchase gap
seen at the end of December means something different for a
Christmas-gift shop than the same gap seen at the end of June. The
remark is never operationalized. No event-rate consequence is derived
or measured, and the remedy offered is a model-based alive-probability
rather than a changed label. For the large part of practice that
consumes discrete labels, the artifact remained unattributed and the
label-level fix undescribed.

\paragraph{Related window artifacts in other fields}
Epidemiology knows a structurally similar problem: in fixed-cohort
perinatal studies the calendar placement of the study window
manufactures artificial seasonality in birth outcomes \citep{strand2011}.
The mechanism differs: truncation of the risk set at the window edges
rather than a two-window comparison. The remedy there is
calendar-aware boundaries through spans covering whole years.
Quasi-experimental design catalogues the generic form of the threat:
a pre--post comparison on a seasonal outcome confounds season with
change, controlled by year-matched comparison periods
\citep{shadish2002}. That literature, however, governs study design on
aggregate series. It does not derive the event-rate function of a
threshold label over an entity panel, measure an artifactual share, or
trace the consequences into a trained classifier.
Within machine learning, seasonal variation of a class prior is label
shift \citep{morenotorres2012}. Label shift, however, is
genuine movement of the prior. Here the prevalence of true
decline can be perfectly stationary while the \emph{measured} rate
swings by a factor of four, because the label itself depends on the
calendar. In the label-noise taxonomy of \citet{frenay2014} this is
noise that is not at random: the mislabeling probability depends on a
covariate, here the anchor's calendar position. The noise is
systematic, cyclic, and shared by every entity labeled at the same
anchor. \citet{nagaraj2024} formalize stochastic time-dependent label
noise. The dependence here is deterministic and construction-induced,
removed by changing the construction. Monitoring practice routinely
builds seasonal baselines into detectors; what those adjust is an
alerting threshold, not a supervised training label.

\section{The construction and the artifact}
\label{sec:construction}

\subsection{Notation and labels}

Let $y_{i,t}\ge 0$ be entity $i$'s activity (spend, units, visitors) in
calendar month $t$. Fix a window length $k$ (we use $k=6$ throughout the main
results, following the production system) and thresholds $\theta\in(0,1)$ for
decay and $\gamma>1$ for growth. At anchor month $a$ define
\begin{align}
B^{\mathrm{adj}}_{i,a} &= \textstyle\sum_{t=a-k+1}^{a} y_{i,t},
&
F_{i,a} &= \textstyle\sum_{t=a+1}^{a+k} y_{i,t},
&
B^{\mathrm{yoy}}_{i,a} &= \textstyle\sum_{t=a+1-12}^{a+k-12} y_{i,t}.
\end{align}
$B^{\mathrm{adj}}$ is the trailing baseline of the standard construction and
$F$ the outcome window. $B^{\mathrm{yoy}}$ is the proposed baseline: the same
$k$ calendar months as the outcome window, one year earlier. The labels are
\begin{equation}
D^{\mathrm{adj}}_{i,a} = \mathbf{1}\{F_{i,a} \le \theta\, B^{\mathrm{adj}}_{i,a}\},
\qquad
D^{\mathrm{yoy}}_{i,a} = \mathbf{1}\{F_{i,a} \le \theta\, B^{\mathrm{yoy}}_{i,a}\},
\end{equation}
each defined only where its baseline is strictly positive, with growth labels
defined symmetrically ($F \ge \gamma B$). Both constructions are leak-free in
the supervised sense. Features computed from months $\le a$ never overlap the
outcome window, and $B^{\mathrm{yoy}}$ lies entirely in the past.

\subsection{Why the adjacent label is calendar-dependent}

The two baselines differ in exactly one respect: $B^{\mathrm{yoy}}$
covers the same points of the seasonal cycle as $F$, while
$B^{\mathrm{adj}}$ covers the $k$ months before them.
For a purely seasonal entity anchored just after its peak, the trailing
baseline sums the high season and the outcome window sums the low
season. The adjacent ratio then reads 0.5, and the label declares
``decay'' with no decline anywhere in the series. The year-over-year
baseline reads parity. Six months later the same entity fires a false
``growth'' event
by the mirrored argument (\ref{app:proof} draws the
three windows). The consequence is easiest to state in the noise-free
case.

\begin{proposition}
\label{prop:mechanism}
Let $y_{i,t} = c_i\, s_{m(t)}$ with $c_i>0$ and a strictly positive seasonal
profile $s_1,\dots,s_{12}$, where $m(t)$ is the calendar month of $t$. Write
$W(j)=\sum_{l=j}^{j+k-1} s_l$ (indices modulo 12) for the twelve $k$-month
window sums, and define the anchor-month ratio
$\rho(m) = \bigl(\sum_{j=1}^{k} s_{m+j}\bigr)\big/\bigl(\sum_{j=0}^{k-1}
s_{m-j}\bigr) = W(m+1)/W(m-k+1)$. Then:
(i) $D^{\mathrm{adj}}_{i,a}=\mathbf{1}\{\rho(m(a))\le\theta\}$ and the
adjacent growth label equals $\mathbf{1}\{\rho(m(a))\ge\gamma\}$: both are
functions of the anchor's calendar month alone, identical for every entity
with the same profile and firing at the same anchors every year;
(ii) $\prod_{m=1}^{12}\rho(m)=1$; hence either $\rho\equiv 1$ or some anchor
months have $\rho(m)<1$ and others $\rho(m)>1$, and $\rho\equiv 1$ holds
exactly when the twelve window sums are all equal --- which occurs for every
profile at $k=12$, and at $k<12$ for profiles whose period divides $k$;
(iii) $D^{\mathrm{yoy}}_{i,a}=0$ at every anchor, and so is the aligned
growth label, since $F_{i,a}=B^{\mathrm{yoy}}_{i,a}$ makes the aligned
ratio identically one and $\theta<1<\gamma$.
\end{proposition}

The proof (a rearrangement argument for the product identity, a
$\gcd$-orbit argument for the equivalence) is in \ref{app:proof}.

Part (ii) is the mechanism in one line. The twelve anchor-month ratios
have geometric mean exactly one. Any profile whose window sums are not
all equal therefore places some anchors below parity. A threshold label
fires annually at those anchors once the shortfall reaches $\theta$. The
growth label fires at the anchors above parity (at $k=6$ these mirror
the decay anchors, since $\rho(m{+}6)=1/\rho(m)$). A sinusoidal
calibration of the noise-free onset is in \ref{app:proof};
in the synthetic panel, noise moves the onset much
earlier. False events appear in volume from peak-to-trough swings of
about 2.3. Shape matters as much as size: a profile whose twelve window
sums are equal generates no events at any amplitude, and at $k=12$ no
profile does. Annual adjacent windows are calendar-neutral by
construction, a remedy measured in Section~\ref{sec:results-remedies}.
The population-level event rate at anchor month $m$ is therefore a
repeating function of the calendar, with a shape that follows the
panel's seasonality. The year-over-year comparison is calendar-neutral
at every $k$. The proposition assumes multiplicative seasonality; under
additive seasonality the dependence persists but attenuates with entity
level. The empirical measurements that follow assume neither form,
since rate curves and disagreement shares are computed on the panels
directly.

Real data adds noise, trends, and entry and exit. The clean dichotomy
of Proposition~\ref{prop:mechanism} then becomes a quantitative
question. The rest of the paper measures it with two population-level
diagnostics any practitioner can compute in a few lines of code. The
first is the \emph{anchor-month rate curve}: the event rate as a
function of the anchor's calendar month, pooled across years, with
entity-cluster bootstrap uncertainty. Where a panel contributes only
one observed year we say so. The instrument is credit scoring's
vintage-analysis plot \citep{siddiqi2006} applied to the
label rather than to the portfolio (\citet{georgiou2022} display a
sequential-slice cousin). What is new is pooling by calendar month and
reading the curve as an audit of the construction. The second is the
\emph{aligned-label disagreement share}: the fraction of adjacent-window
events with no event under the year-over-year definition. On real
panels this is a diagnostic, not a ground-truth error rate;
Section~\ref{sec:results-decomp} decomposes what the disagreements
actually are. Only on synthetic data, where truth exists by
construction, do we call events false positives.

\section{Data and experimental design}
\label{sec:design}

\subsection{Panels}

The originating system was a buyer-health classifier at a
business-to-business service marketplace (named in the declarations; all figures
reported as rates, ratios, and counts). The activity series is monthly
gross transaction value per buyer account. The full panel holds
5{,}712 buyers over 66 months, of whom 2{,}001 pass the same activity
fence used on the public panels. It is measured under the unified
protocol of Section~\ref{sec:protocol}. The deployed system additionally
applied a materiality floor, keeping the 383 buyers large enough for
account managers to act on. The model experiment in
Section~\ref{sec:results-roc} draws its label-defined rows from that
system's training panel of 10{,}426 buyer--month observations. The
material-cohort row of Table~\ref{tab:panels} applies the same floor
inside the unified protocol instead: a buyer qualifies when its final
trailing year clears the floor. This admits 496 entities over 18{,}597
observations. The deployed 383 is the same floor frozen at one training
snapshot. The deployed target was the adjacent-window construction with
$k=6$, $\theta=0.5$ for decay, and $\gamma=1.5$ for growth. Features
were past-only momentum and engagement summaries. The classifier was a
pair of gradient-boosted trees with isotonic calibration on a temporal
holdout. Seasonality is
substantial: 214 of 383 material buyers swing more than 50\% from peak
to trough within a year. That is the regime where the synthetic sweep
shows the adjacent label generating false events in volume.

Three public panels test whether the artifact is a property of the
construction or of one firm's data. Two of them are mechanism panels
rather than customer panels: M5 entities are products and tourism
entities are regions. They replicate the label arithmetic's behavior
on seasonal activity series of any kind. The customer-construct
evidence rests on the production panel and on Online Retail~II, the one
public panel whose entity is a customer. Decision-level claims in this
paper rest on the customer-level evidence only. All are monthly; entity
counts refer to entities contributing at least one valid labeled
observation. \emph{M5 retail} \citep{makridakis2022}: daily Walmart unit
sales aggregated to 64 complete calendar months, labeled at two
granularities. These are the 70 store--department aggregates, and the
2{,}000 highest-volume store--item series among those with at least
90\% positive months after launch. \emph{Australian tourism}
\citep{wickramasuriya2019}: the bottom series of the TourismLarge
hierarchy (regional visitor nights; 274 of 304 input series contribute),
228 months long, the panel with the most statistical power. \emph{UCI
Online Retail~II} \citep{chen2019uci}: per-customer monthly spend for a
UK wholesaler (cleaning steps stated in the replication package). It
holds 5{,}850 customers over 24 complete months, of whom 298 pass the
fence with both labels defined. Its short span limits the joint
comparison to seven anchors in a single year. It shows the artifact on
real customers while the larger panels carry the statistical power.

Two thousand synthetic entities over 72 months isolate the mechanism.
Each has multiplicative month-of-year seasonality with a random peak
month and amplitude $a\in\{0,0.2,0.4,0.6,0.8\}$ (400 entities per
amplitude), lognormal multiplicative noise, and \emph{no trend and no
true decline}. On this clean cohort every decay event is false by
construction. Amplitudes of $a\ge 0.4$ correspond to peak-to-trough
swings above 2.3, the regime most of the production cohort occupies. A
separate seeded 10\% of entities receive a $-60\%$ permanent level shift
at a random month, to measure recall of genuine decline. The $a=0$ cell
is a negative control.

\subsection{Protocol}
\label{sec:protocol}

All panels run one engine that mirrors the production label arithmetic.
An entity--anchor observation is valid when at least 75\% of the
trailing twelve available months are positive (the public analogue of
the production materiality floor). The relevant baseline must also be
positive. The primary comparison set is the \emph{joint} set of
observations on which both labels are defined. Keeping composition
identical means any difference comes from the definitions. The adjacent
label's full valid set is reported secondarily. The first two primary
metrics are the anchor-month rate curve pooled across years and its
dispersion. Dispersion is read as the coefficient of variation
(standard deviation over mean) across the twelve calendar-month rates,
and as the max/min rate ratio. To
these we add the paired dispersion contrast and the aligned-label
disagreement share. Uncertainty comes from a cluster
bootstrap over entities (1{,}000 draws); a sensitivity scheme reported
with the robustness results additionally resamples anchor years.
Sensitivity runs repeat the analysis at $\theta\in\{0.4,0.6\}$ and $k=3$.
The design, pass criteria, and kill criteria for the synthetic and
public replications were fixed in a written plan before any arm
executed. The run artifacts embed code version, seed, and
configuration. The production measurements and the analyses of
Sections~\ref{sec:results-decomp}--\ref{sec:results-roc} were added
afterwards in response to audit questions. The engine, fence, and
metrics were already frozen by that plan. Where an analysis was not
pre-specified we say so.

\section{Results}
\label{sec:results}

\subsection{The anchor-month artifact replicates on public panels}
\label{sec:results-artifact}

\begin{figure}[t]
\centering
\includegraphics[width=0.65\textwidth]{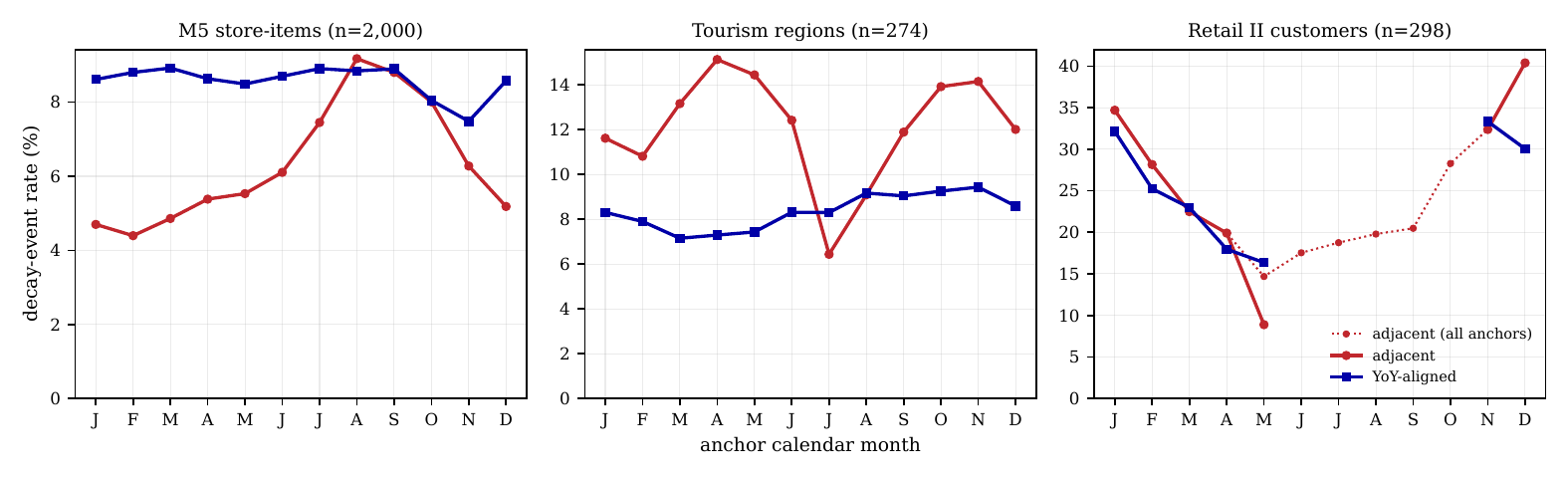}
\caption{Decay-event rate by anchor calendar month, pooled across years,
for the adjacent-window label (red) and the year-over-year-aligned label
(blue), on the joint set where both are defined. The dotted red line in
the right panel is the adjacent label on its full valid set, which
covers all twelve anchor months; its joint set spans seven anchors in a
single observed year. Production panel omitted (proprietary); its
curves are described in the text.}
\label{fig:ratecurves}
\end{figure}

Figure~\ref{fig:ratecurves} shows the anchor-month rate curves and
Table~\ref{tab:panels} summarizes them; every row is produced by the
same engine, fence, and bootstrap. On the production material cohort
(the deployed materiality floor applied to the full panel) the adjacent
decay rate troughs at 3.9\% in April and peaks at 9.1\% in November.
That is a max/min ratio of 2.4, with a coefficient of variation of
0.284. The
aligned label sits nearly flat between 7.1\% and 8.6\% (CV 0.054).
Half of the adjacent events (50\%, interval $[45, 54]$) have no aligned
counterpart. On the full
panel of all 2{,}001 fence-passing buyers the artifact is unambiguous
but milder (ratio 1.3, CV 0.090 against 0.020, disagreement 28\%). The
materiality floor supplies the dose that explains the attenuation.
Raising it from zero through \$50K, \$100K, and \$200K of trailing
annual value moves the ratio $1.3 \to 2.1 \to 2.4 \to 3.4$. The
disagreement share moves $28\% \to 48\% \to 50\% \to 54\%$. Small,
irregular buyers trip a 50\% label constantly for reasons unrelated to
the calendar. The accounts large enough to act on are the smooth,
seasonal ones. The public rows repeat the pattern, with ratios of 2.1
(M5 store--items), 2.4 (tourism), and 4.5 (Retail~II customers). The
Retail~II figure is read under the caveat that this panel contributes a
single observed year.
The aligned label's dispersion sits far below the adjacent label's on
every panel. In every panel the bootstrap exceedance probability $P_b$
that the adjacent label's dispersion exceeds the aligned label's is at
least 0.998. Those draws resample entities with calendar time held
fixed. Section~\ref{sec:results-robust} reports the stricter variant
that also resamples years.

Two details of Figure~\ref{fig:ratecurves} matter. The peaks land where
each panel's own seasonality puts them, with no common macroeconomic
date. The results also bracket the artifact's scope. At high
aggregation it vanishes for lack of events: M5's store--department
level produced \emph{zero} decay events in 3{,}290 valid observations.
At the other extreme noise dilutes it, as the full production panel
shows. The artifact is therefore worst in the middle: substantial
entities with real seasonality, exactly the population decline
prediction is run on.

\begin{table}[t]
\centering
\small
\caption{The artifact and its correction across panels, every row under
the unified protocol of Section~\ref{sec:protocol}. Ratio is the max/min
of the pooled anchor-month rates of the adjacent label on the joint set
(twelve observed anchor months everywhere except Retail~II: seven, in a
single observed year); CV is the coefficient of variation across those
rates; $P_b$ is the entity-cluster bootstrap probability (1{,}000 draws)
that the adjacent label's CV exceeds the aligned label's; Disagr.\ is the
share of adjacent decay events with no aligned event, with 95\%
cluster-bootstrap intervals. Ratios use unrounded rates. Material
cohort: the deployed \$100K materiality floor applied to the full
panel.}
\label{tab:panels}
\setlength{\tabcolsep}{3pt}
\begin{tabular}{lrrrrrr}
\toprule
Panel & Entities & Obs. & Ratio & CV adj/yoy & $P_b$ & Disagr.\ \% \\
\midrule
Prod., all buyers   & 2{,}001 & 51{,}509 & 1.3 & 0.090 / 0.020 & $>$0.999 & 28 [26,29] \\
Prod., material cohort & 496 & 18{,}597 & 2.4 & 0.284 / 0.054 & $>$0.999 & 50 [45,54] \\
M5 store--items          & 2{,}000 & 84{,}742 & 2.1 & 0.249 / 0.047 & $>$0.999 & 37 [34,39] \\
Tourism regions          & 274   & 47{,}503 & 2.4 & 0.195 / 0.090 & $>$0.999 & 69 [65,72] \\
Retail~II customers      & 298   & 1{,}444  & 4.5 & 0.366 / 0.246 & 0.998 & 37 [29,45] \\
M5 store--departments    & 70    & 3{,}290  & \multicolumn{3}{c}{no decay events occur} & --- \\
\bottomrule
\end{tabular}
\end{table}

\subsection{The synthetic sweep isolates the mechanism}
\label{sec:results-synth}

\begin{figure}[t]
\centering
\includegraphics[width=0.65\textwidth]{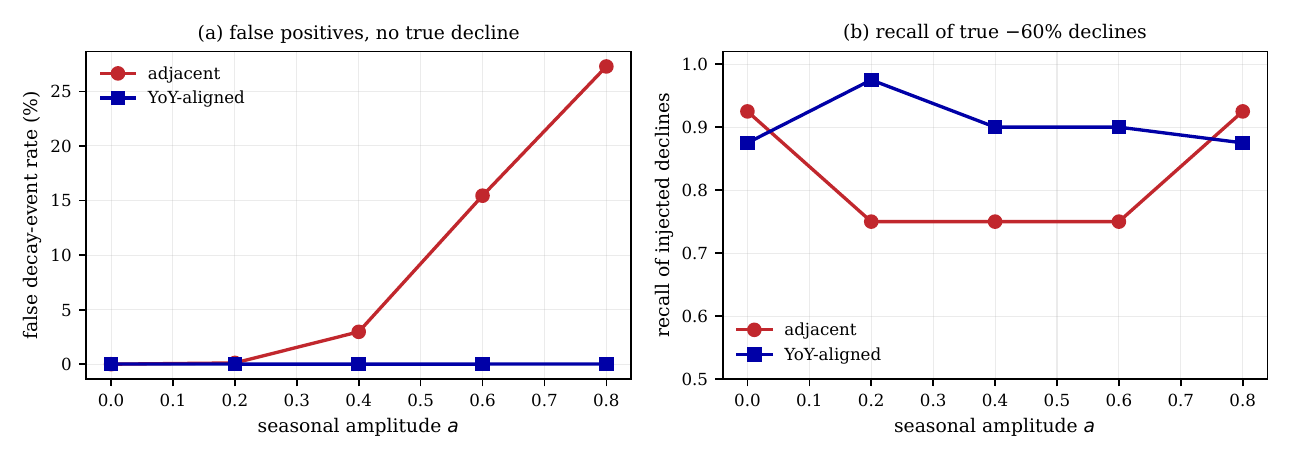}
\caption{Synthetic amplitude sweep, 400 entities per amplitude
(peak-to-trough swing $(1{+}a)/(1{-}a)$), 72 months, $k=6$, $\theta=0.5$.
(a) False decay-event rate on the clean cohort. (b) Recall of injected
$-60\%$ level shifts: at seasonal amplitudes the adjacent label also
\emph{misses} genuine declines.}
\label{fig:dose}
\end{figure}

Figure~\ref{fig:dose} shows the amplitude dose--response on the clean
synthetic cohort, where every decay event is false by construction. The
adjacent label's false-event rate rises from below 0.01\% at $a=0$
through 3.0\% at $a=0.4$ to 27.3\% at $a=0.8$. The aligned label
stays at or below 0.03\% at every amplitude. The $a=0$ negative control
behaves (rates differ by 0.02 percentage points; recall shows no stable
ordering across ten seeds). On the injected true declines, recall is
0.905 aligned against 0.820 adjacent (Figure~\ref{fig:dose}b). At
seasonal amplitudes the adjacent label's recall drops to 0.75: a
decline whose outcome window lands in the high season is masked by the
mismatched baseline. Its apparent recovery at $a=0.8$ comes from
firing on 27\% of all clean observations, catching declines partly by
firing indiscriminately. The adjacent construction not only adds false
positives; it misses genuine declines the aligned comparison catches.

On the real panels the artifact also concentrates where seasonality is
strongest. On M5 store--items the disagreement share is 30--31\% in the
two lower seasonal-strength terciles and 50\% in the top. On tourism it
climbs from 53\% to 88\% (figure, dispersions, and values in
\ref{app:terciles}).

\subsection{Pooling does not repair the labels}
\label{sec:results-pool}

Every rate, share, and interval in Table~\ref{tab:panels} is computed
with all anchor months included. The estimates therefore describe
exactly the pooled regime that multi-slice training
\citep{gurali2014,gattermann2021,georgiou2022}, full-year observation
windows \citep{siddiqi2006}, and overlapping month-shifted sets
\citep{accenture2014patent} produce. Pooling succeeds at what it is for:
the training mixture stops depending on any single anchor's calendar
position. But it cannot change any individual example's stamp. Pooled
across every anchor month, 37--69\%
of the adjacent label's decay events on the public panels have no
aligned counterpart. More than half of the events the aligned label
identifies are absent from the adjacent label (53\% on M5, 55\% on
tourism, 59\% on the production material cohort). Pooling fixes the
composition of the training set without fixing the labels inside it.

\subsection{What the disagreement events actually are}
\label{sec:results-decomp}

Calling a disagreement a false positive assumes the aligned label is
right. There are disagreements where the opposite reading is
plausible. We therefore decompose every adjacent-fires, aligned-silent
event on the production material cohort and on M5 into mechanical
buckets. The buckets are ordered so that readings favoring the adjacent
label are
counted first (definitions and full shares in
\ref{app:decomp}; not pre-specified, reusing the frozen engine and fence). On
production, 578 of the material cohort's 1{,}161 adjacent events are
disagreements. The share is not a pure seasonal-error rate. The
depressed-prior-year bucket, where the adjacent label is most plausibly
right, holds a minority on both panels (15\% production, 24\% M5). The
largest bucket is pullback from recent year-over-year growth. The
two labels answer different questions: decline against the recent run
rate, and decline against the same season last year. Judged without
precedence, however, 83\% of the production disagreements on which the
deseasonalized label is defined are removed by seasonal adjustment alone
(70\% on M5). Whatever else is in those events, the fire needed the
seasonal shape. On entities that decline in two consecutive years --- a
hindsight, label-free criterion --- the direction reverses. The aligned
label fires on
27--41\% of following-year anchors against 6--19\% for the adjacent
label, whose trailing baseline ratchets down with the decline. The
audit also runs in reverse. \ref{app:decomp} decomposes the events the
aligned label \emph{adds}; the largest production bucket, 46\%,
reads decline against a prior-year ramp-up (not pre-specified). The
genuine blind spot is decline followed by \emph{stabilization}, a
current-state question; Section~\ref{sec:practice} describes the
watchlist maintained for it.

\subsection{Alternative remedies, measured}
\label{sec:results-remedies}

Three remedies short of changing the comparison year suggest themselves,
and we run each through the same engine: adjacent labels on a
\emph{deseasonalized} series, with per-entity month-of-year indices
estimated leak-free from months up to the anchor (defined once every
calendar month has been observed twice: 30 months of history); the same
with \emph{pooled} cross-entity indices (18 months); and \emph{annual
windows}, $k=12$, which Proposition~\ref{prop:mechanism}(ii) makes
calendar-neutral by construction (24 months). Table~\ref{tab:remedies}
reports each arm's rate-curve dispersion on its own valid rows, plus
ground-truth error rates on the synthetic clean cohort.

\begin{table}[t]
\centering
\small
\caption{Remedy arms under the unified engine ($k=6$, $\theta=0.5$ unless
stated). History is the months of data needed before the first label.
Synthetic columns are ground truth (clean-cohort false-positive rate;
recall of injected $-60\%$ steps); panel columns are the anchor-month
rate-curve CV on each arm's own valid rows, which is why the adjacent
row's CVs differ slightly from Table~\ref{tab:panels}. The
$k=12$ arm detects a different event (annual decline) over a window twice
as long, which is why its recall exceeds the others'.}
\label{tab:remedies}
\setlength{\tabcolsep}{4pt}
\begin{tabular}{lrrrrrr}
\toprule
Label & History & FP$_{\mathrm{syn}}$ & Recall$_{\mathrm{syn}}$ &
CV$_{\mathrm{M5}}$ & CV$_{\mathrm{tour}}$ & CV$_{\mathrm{prod}}$ \\
\midrule
Adjacent                  & 12 & 9.2\%  & 0.82 & 0.235 & 0.195 & 0.267 \\
Deseason.\ (pooled idx)   & 18 & 9.5\%  & 0.83 & 0.216 & 0.186 & 0.121 \\
Deseason.\ (per-entity)   & 30 & 0.0\%  & 0.90 & 0.158 & 0.084 & 0.107 \\
Annual windows ($k=12$)   & 24 & 0.0\%  & 0.95 & 0.070 & 0.050 & 0.056 \\
Year-over-year aligned    & 18 & 0.0\%  & 0.91 & 0.047 & 0.090 & 0.054 \\
\bottomrule
\end{tabular}
\end{table}

The pattern is consistent. Pooled indices fail where seasonal phases are
heterogeneous --- a panel-level index cannot serve entities that peak in
different months. Per-entity indices work where textbook seasonal
adjustment works, on stationary strictly periodic seasonality (the
synthetic panel) and the long, stable tourism panel. On M5 and
production, where profiles drift and history is shorter, they leave two
to more than three times the aligned label's dispersion. They also
demand a year more history. Annual windows are nearly calendar-flat, as
the proposition predicts, but they detect a different event: annual
decline over a two-year span. That event is blind to a six-month
decline that recovers within
the year. Among the arms that flatten the curve, only the aligned label
keeps the six-month horizon, at tied-lowest history cost.

\subsection{With the model held fixed, relabeling moves holdout skill}
\label{sec:results-roc}

The production remediation provides a controlled comparison. The
classifier pair, feature set, training procedure, and temporal holdout
were held fixed; only the target definition changed, from adjacent to
year-over-year aligned. Table~\ref{tab:roc} reports it. ROC-AUC rises
from 0.767 to 0.864 for the decay direction and from 0.736 to 0.857 for
growth. Precision--recall AUC rises from 0.253 to 0.367 and from 0.430
to 0.743. The caution: the two rows of each pair score different
response variables with different prevalences (decay: 5.5\% adjacent
against 7.0\% aligned at training). The comparison therefore does not
show one model beating another on a fixed task. It shows the aligned
target is learnable where the adjacent target partly is not. A target
that oscillates with the calendar forces the learner to spend its
capacity on seasonality.

The rebuttal is the synthetic panel, where a common referee exists.
Ramp declines are added to the world: a linear fall to $-60\%$ over
twelve months, the decline shape least favorable to an annual
comparison. We train the same learner three times: on the adjacent
label, on the aligned label, and on the adjacent label with
month-of-anchor indicator
features added. All three are scored on identical held-out rows against
the data-generating truth (exact specification in
\ref{app:classifier}). Detecting declines in progress, the aligned-trained model
reaches ROC 0.78. The adjacent-trained model scores 0.44, below chance,
having learned to rank seasonal descent above genuine decline. The
calendar features repair nothing (0.44). Against its own labels the
adjacent-trained model scores 0.96 --- a model can look excellent
against a defective target. Predicting decline \emph{onsets} before any
evidence exists is impossible by construction. There every arm sits
at or modestly above chance at the primary truth cut (ROC 0.49--0.60),
reported as the floor. (This analysis was not pre-specified.)

\begin{table}[t]
\centering
\small
\caption{Production holdout comparison with features, learner, and temporal
holdout held fixed; only the label definition changes.}
\label{tab:roc}
\begin{tabular}{llcc}
\toprule
Direction & Label & ROC-AUC & PR-AUC \\
\midrule
Decay  & adjacent-window        & 0.767 & 0.253 \\
Decay  & year-over-year aligned & 0.864 & 0.367 \\
Growth & adjacent-window        & 0.736 & 0.430 \\
Growth & year-over-year aligned & 0.857 & 0.743 \\
\bottomrule
\end{tabular}
\end{table}

One alternative explanation needs to be ruled out: perhaps the adjacent
target was simply too hard for this model class. It was not. Swapping
learners under the same labels moved holdout skill insignificantly or
negatively. The strongest challenger, the TabPFN tabular foundation
model \citep{hollmann2025}, changed precision--recall AUC by $-0.015$
in both directions. The one-line label change produced the
gains above; the binding constraint was the target, not the model. The
same holds at the synthetic tier, where greater capacity is more
faithful to whatever the label teaches (foundation-model arms
\citep{tabfm2026}, the paired protocol, and the reconciliation in
\ref{app:tabfm}).

\subsection{Robustness}
\label{sec:results-robust}

The artifact does not depend on the particular cut. At $\theta=0.4$
and $\theta=0.6$ the dispersion contrast holds on every panel, with
$P_b\ge 0.98$ on the public panels. Disagreement shares there run
between 33\% and 71\% (49\% on the production material cohort).
Shortening the window to $k=3$ sharpens the contrast on three of four
panels and mildly attenuates the adjacent dispersion on tourism. The
contrast holds everywhere (public-panel $P_b>0.999$; per-cell
dispersion, exceedance, and disagreement values for the public panels
in \ref{app:terciles}). Window length interacts with
profile shape, since the window sums of
Proposition~\ref{prop:mechanism} change with $k$. The growth direction
mirrors the decay results with the calendar phase inverted, as the
mechanism predicts. On the production material cohort the adjacent
growth label's anchor-month ratio is 2.1, and 30\% of its events lack
an aligned counterpart.

The dispersion contrast also survives a stricter treatment of
cross-entity dependence, with one disclosed breach. Entities that share
an anchor year share that year's shocks. We therefore add a two-way
cluster bootstrap in the spirit of the pigeonhole bootstrap
\citep{owen2007}: entities and anchor years resampled independently.
This is a post-hoc
addition whose pass bar (exceedance at least 0.99 on panels with at
least five anchor years) was fixed before the run executed. The scheme
leaves the gap's 95\% interval excluding zero on all four measurable
panels. Exceedance is 0.996 on tourism but 0.983 and 0.981 on the two
production rows. Those two values sit below the run's own bar, and we
report the breach rather than
argue it away. With only five year clusters, probabilities this close
to the bar sit at the edge of what so few clusters can resolve
\citep{cameron2008}. Per-panel
intervals, the discard rule, M5's four-year descriptive status, and the
Retail~II degeneracy are in \ref{app:bootstrap}.

The synthetic conclusions are likewise not a property of one random draw
or of the stationary world that generated them. Across ten seeds the
aligned label's false-event rate stays at or below one quarter of the
adjacent label's in every seed at $a\ge 0.4$. Recall stays within the
pre-specified bound in at least nine of ten. In worlds that break
the generator's assumptions --- decaying amplitude, a migrating seasonal
peak, additive seasonality --- the cure persists in every cell with
seasonal amplitude (cell-level detail in
\ref{app:seeds}). The sweep locates one boundary: sustained compounding
growth, under which the year-old baseline sits below the current level
and the aligned cut becomes effectively stricter. This cost is
quantified with the construction's other prices in
Section~\ref{sec:discussion}.

Finally, the model-level claim is checked against a common referee on
every panel long enough to hold one. The same learner is trained under
each label and scored against a calendar-neutral hindsight truth:
annual windows, whose synthetic false-positive rate
Section~\ref{sec:results-remedies} measures at zero. The learner
trained on the aligned label ranks true declines better. On the public
mechanism panels the truth ROC is 0.754 against 0.631 on M5 items and
0.817 against 0.795 on tourism. On the production panel --- the
customer construct no public panel can supply --- the
aligned-trained model scores 0.764 against the adjacent-trained
0.643 on the material cohort. That closes the different-targets gap of
Section~\ref{sec:results-roc}. Full-panel margins narrow under
small-buyer label noise (\ref{app:classifier}), and the
production arm was not pre-specified. The ordering holds across
ratio cuts and a two-year baseline on all three panels, with one
exception to name. Under a slope-based truth both arms sit near chance
on M5's intermittent series and on production, while the aligned arm
stays clearly ahead on tourism. Protocol,
lifts, net-benefit grids, and the truth-form detail are in
\ref{app:classifier}.

\section{The correction in practice}
\label{sec:practice}

Four practical points came out of adopting the aligned label in a
production system.

\emph{What the correction changed operationally.} The served action list
is the unit of cost: every flagged account consumes account-manager
attention. Relabeling shrank the served ``shrinking'' classification by
a third (119 to 79 accounts). The accounts that left were flags with no
aligned counterpart, dominated by the seasonal-shape events of
Section~\ref{sec:results-decomp}. The account whose all-time-high
``shrinking'' flag had triggered the audit resolved to a growth
classification. The premise that the shorter list
keeps the genuine decliners is checkable where truth exists. On
synthetic ground truth the aligned label's recall of genuine declines is
higher, not lower (0.905 against 0.820,
Section~\ref{sec:results-synth}). A currency translation requires
per-account margins and intervention costs that are confidential. The
capacity effect can be bounded without them. Write $G$ for the
genuine decliners common to both lists, and $r$ for the benefit of
intervening on one, in units of the intervention cost. The adjacent list
spends 119 intervention units per cycle to reach what the aligned list
reaches with 79. That is a saving of 40 units per cycle at any $G$. In
net-benefit terms ($Gr-79$ against $Gr-119$) the advantage is those same
40 units at every $r$. At the boundary case $G=79$ this amounts to
roughly double the net benefit at $r=2$, 14\% more at $r=5$, and 6\%
more at $r=10$. The profit-explicit version of this calculation is the
maximum-profit criterion of \citet{verbeke2012} and its expected form,
the expected maximum profit (EMP) of \citet{verbraken2013}. The
confidential unit economics rule both out here. The same
benefit-to-cost grid evaluated on the public-panel classifiers ships
with the replication materials. The
composition change is the decision-level effect: fewer and truer flags
on the same account-management capacity.

\emph{History and materiality.} The aligned baseline needs the outcome
window's calendar months one year back: roughly $12+k$ months of history
before an entity's first label, against $2k$ for the adjacent
construction. It also needs a baseline fence, so near-zero prior-year
seasons do not create degenerate ratios. In the production panel this
cost no coverage at the operating vintage. On younger customer bases it
would delay labeling of new accounts by up to a year.

\emph{The blind spot, measured.} A pure year-over-year comparison goes
quiet wherever the prior year is itself depressed. For entities
\emph{still} declining the aligned label fires more often
(Section~\ref{sec:results-decomp}). The case alignment truly cannot see
is decline followed by stabilization: a current-state question rather
than a new-decline question. The deployed remedy treated it as one.
Trend and momentum features stayed in the model, and a separate
current-state watchlist tracked realized decline. The blind spot also
has a bounded scoring cost. On the
synthetic ground-truth world, the aligned-trained classifier scores
stabilized decliners at roughly five times its clean-negative level
(mean score 0.11 against 0.02). Counting those entities as negatives
--- they no longer need a decline intervention --- lowers the aligned
model's truth ROC by 0.02 to 0.04. The comparison with the adjacent
model is unchanged.

\emph{The insight outlives the model.} Some weeks after the relabeled
classifier deployed, the organization replaced the machine-learned
buyer-health layer with a transparent rule-based classification for
operational reasons. The replacement's signals are themselves
year-over-year aligned; the calendar alignment survived the model it
was discovered in.

\section{Discussion and limitations}
\label{sec:discussion}

For practitioners the summary is short. If decline labels are built from
threshold-ratio window comparisons on seasonal entities, plot the event
rate by the label's anchor calendar month. Dispersion in that curve is
label noise that looks like class prevalence. Its content can be
audited by scoring the same observations under a year-over-year-aligned
definition and decomposing the disagreements. The diagnostic costs a few
lines of code. Where it fires, aligning the label's baseline to the same
calendar months of the prior year flattens the curve at the source.
The price is extra history, a materiality fence, and a blind spot to
decline-then-stabilization. Where strong growth is sustained, the cut
is also effectively stricter (synthetic recall 0.70 against the
adjacent label's 0.86). Pooling anchors remains good practice for
training-set composition, but it balances the noise rather than
removing it. The other standard moves are, by measurement, not
substitutes either.

Four limitations bound the claims. First, the controlled model-skill
comparison (Table~\ref{tab:roc}) is single-organization. The public
panels support the artifact, the mechanism, and the disagreement
accounting. The synthetic ground-truth experiment supports the
model-level claim. The transfer of the production ROC gains is
demonstrated in one system only. Second, our aligned construction
assumes annual periodicity with a twelve-month lag. Entities with
non-annual cycles, or panels dominated by mobile holidays --- the
calendar effects that survive year-on-year comparison in
official-statistics practice \citep{ess2024} --- need alignment to the
same phase of the relevant cycle. We have not tested that. Third, the
aligned baseline replays history. One anomalous prior-year window (a
promotion spike, an outage) corrupts up to $k$ of the following year's
labels, where an adjacent baseline would self-heal within $k$ months.
Entities are also unlabeled for their first $12+k$ months. That excludes
the early-relationship segment where non-contractual defection
concentrates, a cost young customer bases bear most. Fourth, the
construction we correct
is the threshold-ratio label. Inactivity-based churn definitions carry a
related seasonal confound (seasonal quiet reads as churn) with a
different structure our proposition does not cover. The
latent-attrition lineage \citep{bachmann2021,wunderlich2022} avoids
discrete labels altogether. Our results say nothing against those
choices. They say that where
threshold-ratio window labels are used on seasonal entities, the
calendar belongs in the label's definition and not only in the model's
features. The features-only version of that advice is now
measured and found wanting (Section~\ref{sec:results-roc}).

\section{Conclusion}
\label{sec:conclusion}

The adjacent-window decline label, standard for two decades, compares
windows that sit at different points of the seasonal cycle. Its event
rate is therefore a function of the anchor's calendar month. On one
production and three public panels the dependence is large. A short
geometric-mean argument explains it. Neither the pooling remedies on
record nor calendar features in the model repair it. Aligning the
label's baseline
to the same calendar months one year prior flattens the rate curve on
every real-data panel. In a controlled production comparison the
realigned target
raised holdout ROC-AUC by 0.10 (decay) and 0.12 (growth) with the model
untouched. On synthetic ground truth the aligned-trained classifier
detects declines in progress that the adjacent-trained one ranks below
chance. The target definition is part of the measurement instrument; it
deserves the scrutiny the field already gives its evaluation splits.

\section*{Data and code availability}

Scripts, the pre-specified analysis plan, and timestamped result
artifacts sufficient to reproduce every synthetic and public-panel
number accompany this preprint as arXiv ancillary files. The
production panel is proprietary; its owner consented to the publication
of the aggregate statistics reported here. It appears only as rates,
ratios, and counts --- no production observation, account identity, or
monetary level is disclosed.

\section*{Declaration of generative AI and AI-assisted technologies in the
writing process}

During the preparation of this work, the authors used Claude (Anthropic) to
help organize the structure of the manuscript and to improve the clarity of
the writing. After using this tool, the authors reviewed and edited the
content and take full responsibility for the final content of the
publication.

\section*{Declaration of competing interest}

The authors are employed by Field Nation LLC, which operates the
business-to-business service marketplace whose production system is
analyzed in this paper. Field Nation approved the publication of this
work; the manuscript body refers to the firm generically. The authors
declare no other competing interests.

\section*{Data-owner consent}

Field Nation LLC consented to the publication of the aggregate
statistics (rates, ratios, and counts) derived from its proprietary
production panel.

\section*{Funding}

This research received no external funding.

\section*{CRediT authorship contribution statement}

Md Rezwanul Islam: Conceptualization, Methodology, Software, Formal
analysis, Investigation, Data curation, Writing -- original draft,
Writing -- review \& editing, Visualization.
Wael Mohammed: Conceptualization, Project administration, Resources,
Supervision, Validation, Writing -- review \& editing.

\appendix

\section{Proof of Proposition 1}
\label{app:proof}

(i) and (iii) are immediate from substituting $y_{i,t}=c_i s_{m(t)}$; $c_i$
cancels in both ratios. For (ii), as $m$ runs over the twelve months, both
$m+1$ and $m-k+1$ are shifts and hence run over all twelve residues.
Numerators and denominators therefore traverse the same multiset
$\{W(1),\dots,W(12)\}$, and the product is one by rearrangement. For the
equivalence: if the twelve window sums are equal, every ratio is one.
Conversely, $\rho\equiv 1$ says $W(m+1)=W(m-k+1)$ for every $m$, that is,
$W(j)=W(j+k)$ for every $j$. So $W$ is constant on each orbit of the shift
$j \mapsto j+k$ modulo 12. Let $d=\gcd(k,12)$; each orbit has $L=12/d$
starting points. The $L$ windows launched from one orbit cover each
calendar month exactly $k/d$ times, because month $r$ is covered once for
each offset $i\in\{0,\dots,k-1\}$ with $i\equiv r-j \pmod{d}$ and $d$
divides $k$. Summing the $L$ equal window sums of the orbit therefore
gives $L\,W(j)=(k/d)\sum_m s_m$. So $W(j)=\tfrac{k}{12}\sum_m s_m$ --- the
same constant on every orbit --- and all twelve window sums are equal. For
the period clause: a profile whose period divides $k$ fills every window
with the same number of full cycles, so all windows are equal. Conversely,
equal windows force $s_{j+k}=s_j$ for every $j$ (the difference of
consecutive window sums), so the period divides $k$. At $k=12$ every
window is the full-year sum. \qed

Calibration of the noise-free onset: for a sinusoidal profile
$s_m = 1 + a\sin(2\pi m/12)$ with $k=6$, the $\theta=0.5$ threshold is
crossed at $a \approx 0.54$, a peak-to-trough swing of roughly 3.3.
Noise moves the onset much earlier.

\begin{figure}[h]
\centering
\includegraphics[width=0.85\textwidth]{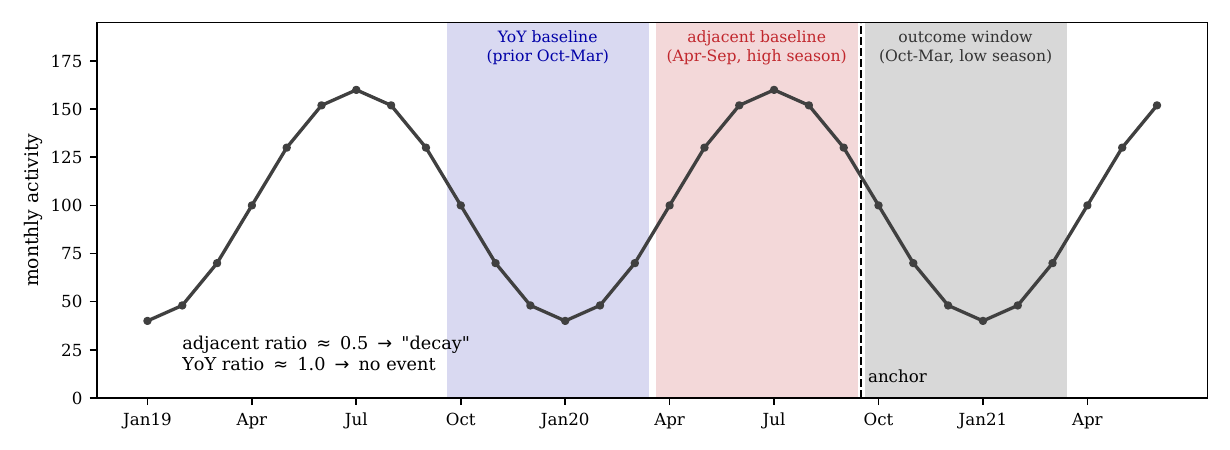}
\caption{The construction and the confound in one picture. A purely
seasonal entity (no trend, no decline) anchored just after its peak: the
adjacent baseline (red span) covers the high season and the outcome
window (grey span) the low season, so the label fires ``decay'' on
seasonality alone; the year-over-year baseline (blue span) covers the
same calendar months one year earlier and reads parity.}
\label{fig:s1schematic}
\end{figure}

Figure~\ref{fig:s1schematic} illustrates the noise-free case behind the
proposition: a purely seasonal entity anchored just after its peak.
There the adjacent baseline covers the high season and the outcome
window covers the low season. The adjacent label therefore fires
``decay'' on seasonality alone. The year-over-year baseline covers the
same calendar months one year earlier and reads parity.

\section{Foundation-model comparison protocol}
\label{app:tabfm}

TabFM \citep{tabfm2026} is a zero-shot in-context tabular foundation model:
it works from an in-context sample rather than a full training pass. The
comparison in the paper is therefore paired at its budget. One seeded,
label-stratified draw of 8{,}192 training rows per label arm serves both as
TabFM's context and as the refit sample for the gradient-boosted baseline.
Both models score the same query set: all 559 in-progress positives and
8{,}000 sampled clean-negative rows. This subsampling is why the
gradient-boosted baseline reads 0.75 in the paired comparison against 0.78
on the full training set. It is also why the tree's own-label 0.95 and
truth 0.42 below differ from the full-set 0.96 and 0.44 reported in the
paper for the same models. Given adjacent-labeled context, TabFM reproduces
its label almost perfectly (ROC 0.96 against the label itself) and ranks
true declines in progress below chance (ROC 0.45). The identical model
given aligned context scores 0.80, ahead of its paired baseline (0.75). The
stronger learner is more faithful to whatever the label teaches (own-label
ROC 0.96 against the tree's 0.95). On the corrected label that fidelity
becomes truth skill. On the adjacent label it becomes a better
seasonality detector while staying below chance on the truth (0.45 against
the tree's 0.42). At the production tier, we swapped the deployed
gradient-boosted pair for the TabPFN tabular foundation model --- an
earlier model than TabFM --- under the same labels. Precision--recall
AUC changed by $-0.015$ in both directions, non-significantly
for decay and significantly worse for growth.

\section{Dependence-aware bootstrap}
\label{app:bootstrap}

The paper's primary uncertainty statements resample entities and hold
calendar time fixed. Entities that share an anchor year share that year's
shocks. Entity resampling alone can therefore understate the variance of
the dispersion gap. This sensitivity was not part of the pre-specified
plan. It was added in response to a methods audit, with its pass bar
fixed before the run executed (exceedance at least 0.99, on panels
observing at least five anchor years).

The scheme is a two-way cluster bootstrap in the spirit of the pigeonhole
bootstrap \citep{owen2007}: per draw, entities and anchor years are
resampled independently with replacement (1{,}000 draws). A draw on which
either label's rate curve is undefined is discarded, leaving at least 999
valid draws on every panel with events. The dispersion gap
$\mathrm{CV}_{\mathrm{adj}}-\mathrm{CV}_{\mathrm{yoy}}$ is then recomputed
from the resampled per-(entity, anchor-year, calendar-month) counts.

Results by panel (two-way scheme): full production panel, 95\% interval
$[0.003, 0.104]$, exceedance 0.983, five anchor-year clusters; production
material cohort, $[0.010, 0.345]$, 0.981, five clusters. M5 items read
$[0.08, 0.24]$, $>$0.999, four clusters (below the bar's own floor, so
descriptive). Tourism reads $[0.02, 0.18]$, 0.996, nineteen clusters.
The two
production exceedance probabilities sit below the run's own 0.99 bar.
The paper reports the breach rather than arguing it away. With five
clusters, probabilities this close to the bar are at the edge of what so
few clusters can resolve \citep{cameron2008}. The honest summary: the
gap's sign is robust to year clustering everywhere it is measurable.
The near-certainty of the entity-only $>$0.999 is not available for
the production rows under the stricter scheme. Retail~II spans only two
anchor years, so the year-clustered scheme is degenerate there and its
interval crosses zero ($[-0.02, 0.22]$). That is the panel's
single-observed-year caveat made quantitative, not evidence against its
curve.

\section{Multi-seed and drift-world sweep}
\label{app:seeds}

Across ten seeds of the synthetic sweep, the aligned label's false-event
rate stays at or below one quarter of the adjacent label's in every seed at
$a\ge 0.4$. Recall of injected declines under the aligned label stays
at or above nine tenths of the adjacent label's in at least nine of ten
seeds. In worlds that break the generator's assumptions the cure persists.
With the seasonal amplitude decaying linearly to a quarter of its starting
value over the panel: aligned false-event rate 0.00005 against adjacent
0.083 at $a=0.8$. With the seasonal peak migrating four months across the
panel: the adjacent artifact worsens (0.287 at $a=0.8$) while alignment
still cures it (0.0023), with comparable recall. With additive rather than
multiplicative seasonality (a common absolute amplitude and arithmetic
noise, clipped at zero): both labels share a noise-floor false-event rate
of about 0.019 at $a=0$. The cure holds at $a\ge 0.4$ (0.162 against
0.012 at $a=0.8$; the factor-of-five cell at $a=0.4$ is the weakest in the
sweep). In no cell with seasonal amplitude ($a\ge 0.4$) does the median
ordering reverse. At $a=0$ under sustained decline the aligned label fires
slightly more often than the adjacent one (median rates of 6.5 against 1.9
events per ten thousand rows). That is the zero-amplitude mirror of the
growth-side trade. The boundary the sweep locates is sustained
multiplicative trend. Under compounding growth of one percent per month,
the year-old baseline sits below the current level and the aligned cut
becomes effectively stricter. Recall of injected declines then drops to
0.70 against the adjacent label's 0.86 at $a=0.8$ while false events stay
near zero. Sustained decline mirrors this favourably (aligned recall
0.97).

\section{Classifier experiments: specification and public-panel results}
\label{app:classifier}

Shared model specification (the paper's synthetic three-arm experiment
and the public-panel experiment below): one past-only feature set --- log
trailing means over one, three, six, and twelve months; three- and
six-month momentum ratios and the trailing year against the prior year;
twelve-month volatility and positivity share; no month-of-year features
--- and one gradient-boosted classifier (300 trees, learning rate 0.05,
31 leaves, fixed seed). In the synthetic experiment the temporal split is
training anchors 24--47 and test anchors 48--65. The third arm appends a
twelve-level month-of-anchor indicator. The decline-onset floor spans
ROC 0.49--0.60 across arms at the primary truth cut.

Public-panel design: each label arm is trained on its own label's valid
rows in the training window (anchors 23 up to the split). Both arms are
scored on a common truth-valid test set in the held-out final anchors
(the last 9 anchor months on M5, the last 24 on tourism). The hindsight
truth is
calendar-neutral: annual windows (next-twelve-month sum at or below half
the trailing-twelve-month sum). The paper's remedy measurement puts that
construction's anchor-month artifact at zero. Results: ROC 0.754 (aligned)
against 0.631 (adjacent) on M5 items and 0.817 against 0.795 on tourism.
Average precision is 0.157 against 0.079 on M5 and 0.288 against 0.180 on
tourism. Lift at the top decile is 3.9 against 2.7 (M5) and 4.6 against
4.0 (tourism). Top-decile net benefit per targeted entity is higher for
the aligned arm at every benefit-to-cost ratio in $\{2, 5, 10\}$.
Own-label
ROC mirrors the production direction on M5 (0.824 aligned against 0.757
adjacent). On tourism the aligned label is harder to fit (own-label 0.864
against 0.899), yet its model tracks truth better. The adjacent model's
own-label skill partly measures learned seasonality. Truth-form
sensitivity: the ordering holds under ratio cuts 0.4 and 0.6 and under a
two-year baseline on both panels. Under a slope-based truth both arms sit
near chance on M5's intermittent series (0.50 against 0.53), while the
aligned arm stays ahead on tourism (0.624 against 0.459). Online
Retail~II cannot support the experiment structurally. Its 24 usable
months cannot hold 24 months of feature history plus a 12-month truth
horizon. That is the role the longer production panel plays.

Production truth-referee arm (added in response to a pre-submission
audit; not pre-specified): the identical design --- same features,
learner, seed, label arms, and annual-window truth --- runs on the
production panel. The panel spans 66 complete months, with training
anchors 23 up to the split and the last 10 truth-evaluable anchors held
out. On the material
cohort (the deployed materiality floor of the paper's Section~4.1) the
aligned-trained model's truth ROC is 0.764 against the
adjacent-trained 0.643. Average precision is 0.101 against 0.088. Lift
at the top decile is 3.65 against 3.11. Top-decile net benefit is
higher for the aligned arm at every benefit-to-cost ratio in
$\{2, 5, 10\}$.
On the full panel the margins narrow (ROC 0.747 against 0.739), the
dilution direction of the paper's Section~5.1. Truth-form sensitivity
mirrors the public panels. The aligned arm wins under ratio cuts 0.4
(0.815 against 0.687) and 0.6 (0.756 against 0.614), and under the
two-year baseline (0.742 against 0.491). Under the slope-based
truth both arms sit near chance (aligned 0.495, adjacent 0.534) ---
the intermittent-series behavior of M5 repeated at the customer level.
Own-label ROC mirrors the direction of the paper's Table~3 (0.874
aligned against 0.775 adjacent on the material cohort).

\section{Seasonal-strength terciles; threshold and window sensitivity}
\label{app:terciles}

\begin{figure}[h]
\centering
\includegraphics[width=0.72\textwidth]{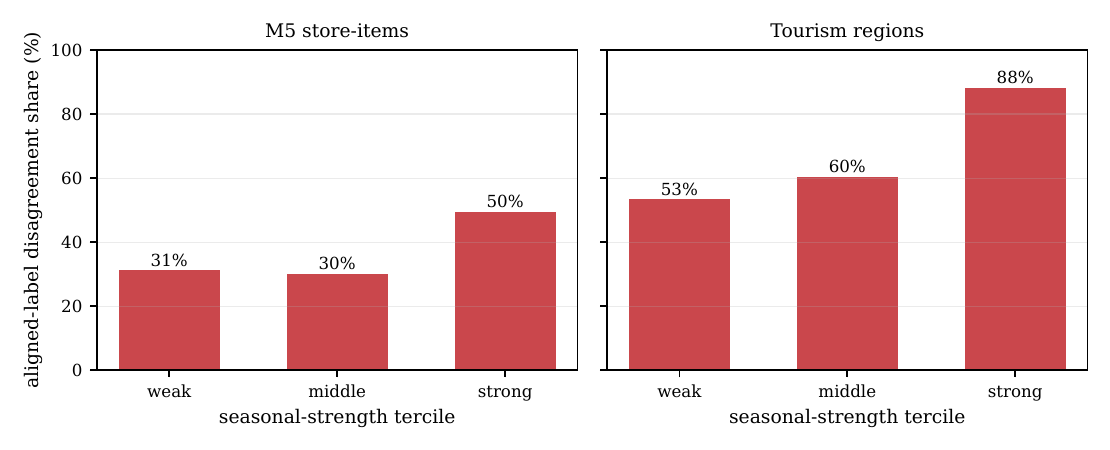}
\caption{Aligned-label disagreement share of adjacent-window decay
events, by seasonal-strength tercile of the entity.}
\label{fig:s6terciles}
\end{figure}

Entities are sorted by a month-of-year $R^2$ measure of seasonal
strength on detrended log activity. The aligned-label disagreement
share of adjacent-window decay events by tercile is then 31\%, 30\%,
and 50\% on M5 store--items and 53\%, 60\%, and 88\% on tourism.
Tercile sizes are 667/666/667 store--items and 102/101/101 tourism
input series. The
adjacent label's anchor-month dispersion rises in step (CV 0.18 to 0.52
on M5, 0.12 to 0.46 on tourism). The aligned label's sits below it in
every tercile. Figure~\ref{fig:s6terciles} shows the shares.

Table~\ref{tab:s6sens} gives the public-panel threshold and window
sensitivity cells. Each cell reports the adjacent/aligned anchor-month
rate-curve coefficients of variation, the entity-cluster bootstrap
exceedance probability $P_b$, and the aligned-label disagreement share
with its 95\% interval. On the production material cohort the CV pairs
are 0.30/0.08 at $\theta=0.4$, 0.30/0.05 at $\theta=0.6$, and 0.34/0.06
at $k=3$. On tourism the $k=3$ adjacent dispersion attenuates from
0.195 to 0.16 while the contrast still holds.

\begin{table}[h]
\centering
\small
\caption{Public-panel sensitivity cells (the main result is $k=6$,
$\theta=0.5$; each row varies one element). CV pairs are
adjacent/aligned; $P_b$ is the entity-cluster bootstrap probability
(1{,}000 draws) that the adjacent CV exceeds the aligned CV; Disagr.\ is
the share of adjacent decay events with no aligned counterpart, with
95\% cluster-bootstrap intervals.}
\label{tab:s6sens}
\begin{tabular}{llrrr}
\toprule
Panel & Cell & CV adj/yoy & $P_b$ & Disagr.\ \% \\
\midrule
M5 store--items & $\theta=0.4$ & 0.25 / 0.08 & $>$0.999 & 33 [30, 35] \\
M5 store--items & $\theta=0.6$ & 0.24 / 0.03 & $>$0.999 & 38 [36, 39] \\
M5 store--items & $k=3$        & 0.33 / 0.05 & $>$0.999 & 37 [35, 39] \\
Tourism regions & $\theta=0.4$ & 0.22 / 0.09 & $>$0.999 & 71 [68, 75] \\
Tourism regions & $\theta=0.6$ & 0.20 / 0.07 & $>$0.999 & 66 [63, 69] \\
Tourism regions & $k=3$        & 0.16 / 0.08 & $>$0.999 & 60 [57, 62] \\
Retail~II customers & $\theta=0.4$ & 0.38 / 0.27 & 0.985 & 35 [27, 44] \\
Retail~II customers & $\theta=0.6$ & 0.33 / 0.21 & 0.999 & 34 [28, 41] \\
Retail~II customers & $k=3$        & 0.58 / 0.27 & $>$0.999 & 37 [32, 43] \\
\bottomrule
\end{tabular}
\end{table}

\section{Decomposition of the aligned-silent disagreements}
\label{app:decomp}

The paper's disagreement decomposition classifies every adjacent-fires,
aligned-silent event on the production material cohort and on M5
store--items into four mechanical buckets. The buckets are ordered so
that readings
favoring the adjacent label are counted first: (i) \emph{depressed prior
year} --- the aligned baseline sits at or below 70\% of its own level one
year earlier, so the aligned silence may itself reflect a bad prior year
and the adjacent reading is plausibly right; (ii) \emph{trailing spike}
--- the trailing baseline is at least 30\% above its level one year
earlier, so the fire is a pullback from recent year-over-year growth;
(iii) \emph{seasonally resolved} --- the leak-free deseasonalized label
of the paper's remedy comparison is defined and silent; (iv)
\emph{unresolved}. Bucket shares under this precedence: production 15\%,
56\%, 13\%, 17\%; M5 24\%, 37\%, 19\%, 19\% (shares rounded
independently, so they need not sum to 100). The buckets overlap
heavily. Judged without precedence, 83\% of the production disagreements
on which the deseasonalized label is defined (345 of 417) are removed by
seasonal adjustment alone, 70\% on M5.

The multi-year check uses a label-free criterion: calendar-year
totals down at least 30\% in each of two successive years. Under it
the aligned label fires on 27--41\% of the following-year anchors
against 6--19\% for the adjacent label (production: 5--8 such buyers per
vintage; M5: 27--63 items). The reason: a steadily declining
entity drags its own trailing baseline down with it. The adjacent ratio
hovers near parity while the aligned comparison still registers the
year-over-year loss.

The mirror decomposition --- every yoy-fires, adjacent-silent event,
the events the aligned label \emph{adds} --- closes the audit's other
direction (added in response to a pre-submission audit; not
pre-specified). The added share is 59\% of aligned events on the
production material cohort (822 of 1{,}405) and 53\% on M5
store--items (3{,}847 of 7{,}264). The buckets reuse the 0.7/1.3
thresholds above, with aligned-unfavorable readings counted first ---
the same convention as the main decomposition, mirrored: (i)
\emph{prior-year spike} --- the aligned baseline at least 30\% above
its own level one year earlier, so the fire reads decline against a
prior-year ramp-up, the replayed-history cost of the paper's
limitations quantified --- 46\% production, 29\% M5; (ii)
\emph{trailing ratchet} --- the trailing baseline at or below 70\% of
its level one year earlier, the multi-year mechanism above --- 28\%
production, 39\% M5; (iii) \emph{seasonally masked} --- the leak-free
deseasonalized label fires, so the raw adjacent silence was itself the
seasonal shape hiding a decline --- 3\% production, 6\% M5; (iv)
\emph{unresolved} --- 23\% production, 26\% M5.

\bibliographystyle{elsarticle-harv}
\bibliography{references}

\end{document}